\RequirePackage{lineno}
\documentclass[aps,pre,twocolumn,groupedaddress,showpacs]{revtex4}
\usepackage{amssymb}
\usepackage{graphicx,color}
\usepackage{textcomp}
\usepackage{amssymb}
\usepackage{placeins}

\definecolor{r}{rgb}{1,0,0}
\definecolor{g}{rgb}{0,1,0}
\definecolor{b}{rgb}{0,0,1}
\begin{document}
%\linenumbers
% Use the \preprint command to place your local institutional report
% number in the upper righthand corner of the title page in preprint mode.
% Multiple \preprint commands are allowed.
% Use the 'preprintnumbers' class option to override journal defaults
% to display numbers if necessary
%\preprint{}

%Title of paper
\title{Diagnosing hyperuniformity in two-dimensional disordered jammed-packings of soft spheres}

\renewcommand*{\thefootnote}{\fnsymbol{footnote}}
\author{Remi Dreyfus$^{1}$, Ye Xu$^{1, 2}$, Tim Still$^{2}$, Lawrence A. Hough$^{1}$, A. G. Yodh$^2$, and Salvatore Torquato$^3$\footnote{Corresponding authors: torquato@princeton.edu; remi.dreyfus@gmail.com} }

\affiliation{$^{1}$Complex Assemblies of Soft Matter, CNRS-Rhodia-UPenn UMI 3254, Bristol, PA 19007-3624, USA}
\affiliation{$^{2}$Department of Physics and Astronomy, University of Pennsylvania, Philadelphia, PA 19104-6396, USA}
\affiliation{$^{3}$Department of Chemistry, Department of Physics, Princeton Institute for the Science and Technology of Materials, and Program in Applied and Computational Mathematics, Princeton University, Princeton, New Jersey 08544, USA}

\date{\today}

\begin{abstract}
Hyperuniformity characterizes a state of matter for which density fluctuations diminish towards zero at the largest length scales. However, the task of determining whether or not an experimental system is hyperuniform is experimentally challenging due to finite-resolution, noise and sample-size effects that influence characterization measurements. Here we explore these issues, employing video optical microscopy to study hyperuniformity phenomena in disordered two-dimensional jammed packings of soft spheres.  Using a combination of experiment and simulation we characterize the detrimental effects of particle polydispersity, image noise, and finite-size effects on the assignment of hyperuniformity, and we develop a methodology that permits improved diagnosis of hyperuniformity from real-space measurements. The key to this improvement is a simple packing reconstruction algorithm that incorporates particle polydispersity to minimize free volume. In addition, simulations show that hyperuniformity can be ascertained more accurately in direct space than in reciprocal space as a result of finite sample-size. Finally, experimental colloidal packings of soft polymeric spheres are shown to be hyperuniform.
\end{abstract}

% insert suggested PACS numbers in braces on next line

\maketitle

%--------------------------------------------------------------------------------------------------

\section{Introduction}

Hyperuniformity is a state of matter characterized by vanishing density fluctuations at large length scales ~\cite{Torquato03}, and the concept of hyperuniformity has emerged as a new way to classify crystals, quasi-crystals and special disordered systems ~\cite{Torquato03,Za09}. Disordered hyperuniform materials, for example, behave more like crystals in the manner they suppress density fluctuations over large length scales, and yet they also resemble traditional liquids and glasses with statistically isotropic structures with no Bragg peaks. During the last decade, hyperuniform disordered states have been identified in maximally random jammed packings of hard particles \cite{Do05d,Za11a,Ji11},  jammed athermal granular systems~\cite{Berthier11}, jammed thermal colloidal packings~\cite{Kurita10}, cold atoms~\cite{Le14}, certain Coulombic systems~\cite{To08c}, and "stealthy" disordered classical ground states~\cite{Ba08}. Furthermore, the hyperuniformity property has been suggested to endow materials with novel physical properties potentially important for applications in photonics~\cite{florescu09, Fl13, man13} and electronics \cite{Wi12, He13,xie13} . Thus, the ability to ascertain the degree of hyperuniformity in a disordered system is becoming increasingly important. 

Colloids offer a potentially fruitful experimental system for design, fabrication and testing of hyperuniform systems. In fact, it has been conjectured that certain jammed colloidal particle packings are in the hyperuniform state ~\cite{Torquato03}.  If true, a significant laboratory challenge would be to learn how one might self-assemble colloids into stealthy~\cite{Ba08} hyperuniform states.  Unfortunately, the task of ascertaining whether or not a colloidal packing is hyperuniform is challenging , and previous  work of Berthier~\cite{Berthier11} and Kurita~\cite{Kurita11}  have demonstrated that this  task is highly non-trivial, because typical experimental data sets have imperfections due to optical imaging and particle-size polydispersity. Thus, we currently lack a general method to test for hyperuniformity in experimental colloidal packings recorded via standard real-space methods such as optical microscopy.

To this end, we employ video optical microscopy to study hyperuniformity phenomena in two-dimensional (2D) disordered, jammed-packings of soft spheres.  Using a combination of experiment and numerical simulation we characterize the detrimental effects of particle polydispersity, image noise, and finite-size effects on the assignment of hyperuniformity.  In typical optical microscopy experiments with micrometer-sized particles, particle centers can be measured with tens-of-nanometer precision, but the particle boundaries are not well resolved. In a related vein, accurate determination of an individual particle diameter is very challenging in the presence of size polydispersity, a characteristic of virtually all real colloidal materials.

Herein, we report on a general method for diagnosis of hyperuniformity of jammed systems of spheres. Guided by computer simulations, we experimentally investigate the effect of polydispersity on the detection of hyperuniformity, and we develop an algorithm that recovers individual particle sizes from digitized images of jammed packings of polydisperse spheres by systematically minimizing free volume and overlaps between particles. The methodology enables us to determine whether a 2D packing of spheres is hyperuniform or not.  We further show that diagnosis of hyperuniformity in finite-size experimental packings using real-space measurements is more accurate than reciprocal-space measurements. We test the method on simulated ideal hyperuniform packings, and then, apply this knowledge to experimental images of 2D soft microsphere jammed-packings. The experimental packings exhibit clear signatures of hyperuniformity which, importantly, are very sensitive to the correct assignment of size polydispersity \cite{Kurita11, Berthier11}.

%---------------------------------------------------------------------------------------------------------------------

\section{Hyperuniformity}

\subsection{Point Configurations}

The hyperuniformity concept applies to many-particle systems with number density $\rho$ in $d$-dimensional Euclidean space $\mathbb{R}^d$ and is intimately related to the suppression of local density fluctuations at long length scales. In order to understand the concept more precisely, consider random placements of a spherical observation window $\Omega$ of radius $R$ (e.g., a circle in $d=2$ or a sphere in $d=3$) in a point pattern (e.g., a lattice of points or centers of 
the atoms of a liquid). The number of points, $N(R)$, contained in $\Omega$ is a random variable, and we define the {\it number variance} $\sigma^2_N(R)$ in an observation window
of radius $R$ to be 
\begin{equation}
\sigma^2_N(R) \equiv \langle N(R)\rangle^2 - \langle N(R) \rangle^2,
\label{N}
\end{equation}
where angular brackets denote ensemble average. For Poisson distributions of points (ideal gas), and for traditional correlated disordered systems such as liquids and glasses, the number variance for large $R$ scales as the window volume, i.e.,  $\sigma^2_N(R) \sim R^d$.

A {\it hyperuniform} point pattern is one wherein the number variance $\sigma_N(R)$ grows more slowly than the window volume for large R, i.e., more slowly than $R^d$ \cite{Torquato03}. Perfect crystals and quasicrystals are hyperuniform with scaling $\sigma^2_N(R) \sim R^{d-1}$, i.e, the variance grows as the window surface area). In disordered systems however, when the variance scales more slowly than $R^d$, then the system should be considered exotic. Hyperuniform states of {\it disordered matter} have hidden order on long length scales that is not apparent at short length scales.

The hyperuniformity condition described above is equivalent to the following condition on the structure factor, $S({\bf k})$, at wave vector $\bf k$ \cite{Torquato03}:
\begin{equation}
\label{S} \lim_{|{\bf k}| \rightarrow 0}S({\bf k}) = 0.
\end{equation}
This condition implies that the infinite-wavelength density fluctuations of the system vanish (when appropriately scaled). The ensemble-averaged structure factor of a point pattern
in $\mathbb{R}^d$ at number density $\rho$ is defined via
\begin{equation}
\label{eq_Sk} S({\bf k})= 1+ \rho {\tilde h}({\bf k}),
\end{equation}
where ${\tilde h}({\bf k})$ is the Fourier transform of the total
correlation function $h({\bf r}) = g_2({\bf r})-1$, and $g_2({\bf
r})$ is the pair correlation function that characterizes the system. (Note, definition (\ref{eq_Sk}) implies that the forward scattering contribution is omitted.) 
For computational purposes, the structure factor, $S({\bf k})$, for
a finite-size point configuration in a fundamental cell under periodic boundary
conditions, can be obtained directly from
the particle positions ${\bf r}_j$ \cite{Za09}, i.e.,
\begin{equation}
S({\bf k}) = \frac{1}{N} \left |{\sum_{j=1}^N \exp(i {\bf k} \cdot
{\bf r}_j)}\right |^2  \qquad ({\bf k} \neq {\bf 0}).
\label{coll}
\end{equation}
Here $N$ is the total number points in the fundamental cell. Note that to ascertain accurately the very small wavenumber behavior of $S(k)$ (crucial for the hyperuniformity test), $N$ (system size) must be large.

\subsection{Multiphase Media}

The hyperuniformity concept has been extended to statistically homogeneous systems made of $2$ (or more) phases in which each phase $i$ ($i=1,2$) occupies volume fraction $\phi_i$.  In the specific case of polymeric particles suspended in water, phase 1 is the polymeric particles while phase 2 is the surrounding aqueous phase. The following is  a general formulation of hyperuniformity; it can be apply whether the system is composed of particles or not. In two-phase systems (e.g., composites, binary suspensions, block copolymers, etc.) \cite{To02a,Sa03,Br06,Zo08}, hyperuniformity is manifested in the suppression of local volume-fraction fluctuations: 
\begin{equation}
\sigma_\tau^2(R) =\left\langle  \tau(R)^2 \right\rangle - \left\langle  \tau(R) \right\rangle^2.
\label{tau}
\end{equation}
Here the random variable $\tau$ is the local volume fraction of either phase in some observation window of radius $R$ \cite{Za09} and hence the average of $\tau$
is simply the volume fraction of the phase of interest (either $\phi_1$ or $\phi_2$). 
Non-hyperuniform disordered media, such as typical liquids and glasses, have the scaling 
$\sigma_\tau^2(R) \sim R^{-d}$. By contrast, volume-fraction fluctuations
in hyperuniform two-phase systems will decrease faster than $R^{-d}$ \cite{Za09}.

It has been shown that this hyperuniformity condition for two-phase
media is equivalent to the following condition on the spectral density  ${\tilde  \chi}({\bf k})$ (defined below)  \cite{Za09}:
\begin{equation}
\lim_{|{\bf k}| \rightarrow 0} {\tilde \chi}({\bf k}) = 0.
\label{chi}
\end{equation}
Here the spectral density, ${\tilde \chi}({\bf k})$, is the Fourier transform of
the autocovariance function 
\begin{equation}
\chi({\bf r})= S_2({\bf r}) -\phi_i^2,
\end{equation}
and $S_2({\bf r}) $ is the two-point probability function for phase $i$: the probability
that the end points  of a vector $\bf r$ lie in phase $i$ when the vector is randomly placed into the medium. Note that 
the spectral density ${\tilde \chi}({\bf k})$ is obtainable  directly  from  scattering experiments  \cite{Debye49}.

For computational purposes, the spectral density, ${\tilde \chi}({\bf k})$, i.e., for a given finite two-phase configuration in a fundamental cell of volume $V$ under periodic boundary conditions, can be obtained directly from the square of the Fourier transform of the configuration \cite{To99}. In particular,
\begin{equation}
{\tilde \chi}({\bf k})= \frac{1}{V} |{\tilde J}({\bf k})|^2,
\label{FT}
\end{equation}
where ${\tilde J}({\bf k})$ is the Fourier transform of $J({\bf x})=I({\bf x})-\phi_i$,
and $I(\bf x)$ is the indicator function for phase $i$ defined by
\begin{equation}
I({\bf x}) =\Bigg\{{1, \quad {\bf x} \in \text{phase}\; i,\atop{0, \quad \text{otherwise}}}.
\end{equation}
\subsection{Detecting Hyperuniformity in Polydisperse Sphere Packings}

Determination of whether a monodisperse packing of spheres is hyperuniform can be established by analyzing the centers of the spheres using the point configuration formulation [Eq. (\ref{N}) or Eq. (\ref{S})], or the space occupied by the spheres using the volume-fraction formulation [Eq. (\ref{tau}) or Eq. (\ref{chi})]. Both methodologies are equivalent. Importantly, if one desires to ascertain whether a sphere packing with {\it polydisperse } size distribution is hyperuniform, then one cannot use the point configuration formulation. In the polydisperse case, the point configuration formulation may lead to a false conclusion that the packing is not hyperuniform due to size disparity. The points in polydisperse samples should be weighted according to the volume occupied by the spheres. In such instances, it has been shown that hyperuniformity can only be inferred from the degree to which local volume-fraction or area-fraction fluctuations are suppressed, as described in Ref. \cite{Zachary11}.

To illustrate the two different local-volume fraction detection methods, consider a maximally random jammed (MRJ) packing of binary disks under the ``strict" jamming constraint; these samples were generated using the numerical algorithm described in Ref. \cite{Zachary11}. It has been established that MRJ packings of particles of general shape and polydisperse size distribution are disordered and hyperuniform \cite{Do05d,Zachary11,Za11a,Berthier11,Kurita11,Ji11}. Thus, the MRJ system provides useful samples for tests of the two hyperuniformity analysis methods in the multiphase formulation. These methods may employ both the direct-space relation for the local-volume fraction variance (Eq.~\ref{tau}) and the spectral (reciprocal-space) density (Eq.~\ref{FT}). In order to make contact with digital images obtained from our experiments, we digitize or binarize the computer-generated jammed binary disk configurations. Therefore, except for digitization, no other uncertainties are introduced. Of course, additional errors will be introduced when the image is extracted from experiment.

\begin{figure}
\includegraphics[width=1.00\linewidth]{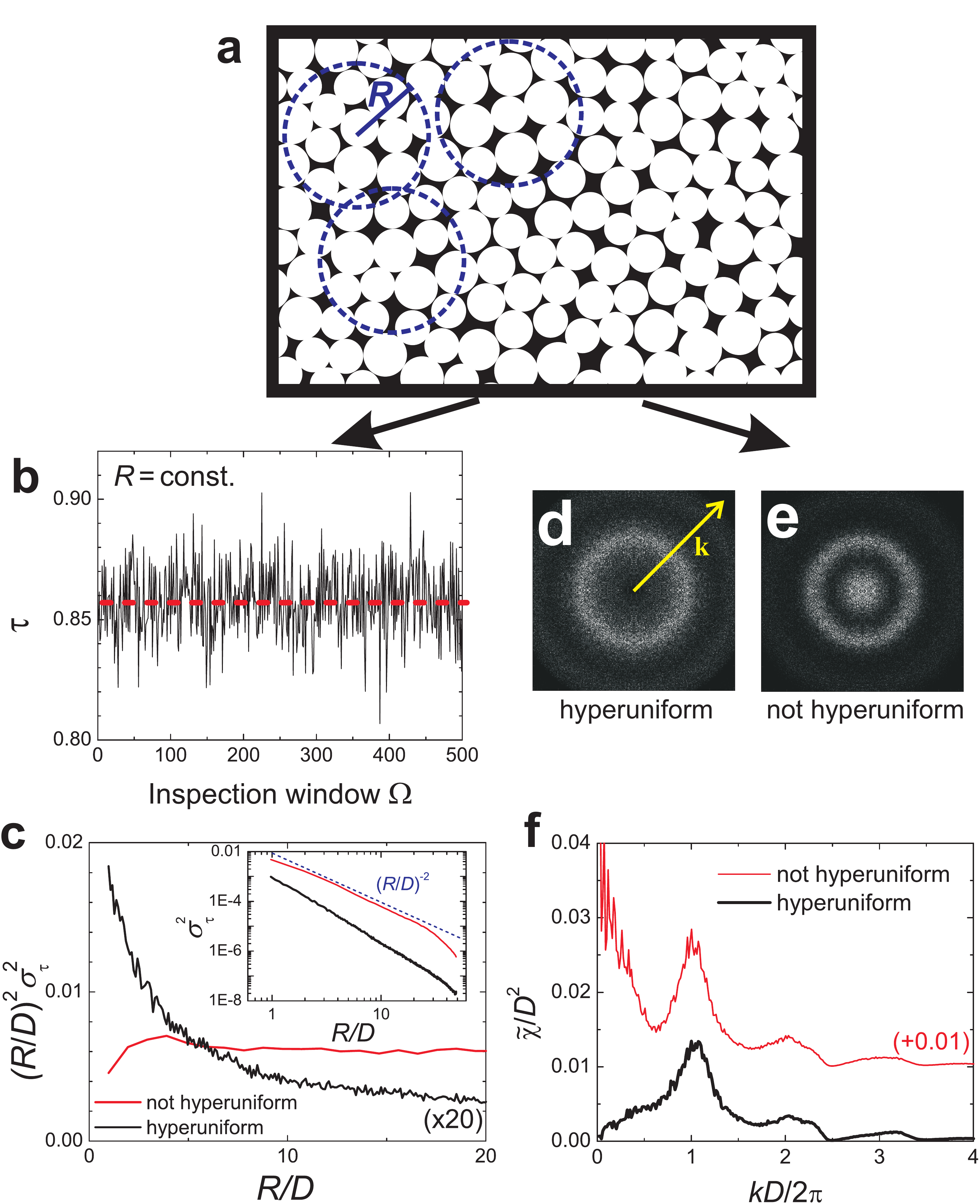}
\caption{(Color online)
 (a) A portion of digitized  (binarized) image of a MRJ 2D packing of 10,000 binary hard disks  (white circles) in which the fraction of large disks is 0.75 and the size ratio is 1.32. Dashed circles indicate some sample circular observation
windows of radius $R$.
(b) Local area fraction of white particles, $\tau$, in 500 independent trial window placements. The dotted line is the packing fraction $\phi_{MRJ}\approx0.86$.
(c) The local-volume fraction variance $\left(R/D\right)^2\sigma^2_\tau(R)$ as a function of $R$ (normalized by the average diameter $D$) for hyperuniform (thick black line) and non-hyperuniform (thin red line) packings. Inset: $\sigma^2_\tau(R)$ as a function of $R/D$  for hyperuniform (thick black line) and non-hyperuniform (thin red line) packings. The dashed curve, included as a guide to the eye, behaves like $(R/D)^{-2}$. Typical power spectra for  hyperuniform (d) and non-hyperuniform (e) packings. 
The arrow shows ${\bf k}$ in one specific direction.
(f) The spectral density $\tilde{\chi}/D^2$ versus dimensionless wavenumber $k D/(2\pi)$ as obtained by angular averaging of (d) and (e).}
 \label{fig1}
\end{figure}

Figure~\ref{fig1} summarizes the real-space and reciprocal-space approaches to ascertain hyperuniformity based on ideal binarized images. A portion of such an image is shown in Fig.~\ref{fig1}a. Particles appear bright and the background appears dark. 
Fig.~\ref{fig1}b and c summarize the analysis in real-space. A series of circular observation windows of radius $R$ is used to randomly sample the binarized image, which is made dimensionless by the average diameter $D$, and the local area fraction occupied by the disks is recorded for each observation circle. Fig.~\ref{fig1}b shows  the local area fraction $\tau$ for 500 different observation windows at one particular $R$. In this specific example, $\tau$ fluctuates around a mean value close to the overall packing-fraction ($\phi_{MRJ}\approx0.86$, dotted line). We calculate the 
variance $\sigma_\tau^2$ for this value of $R$, and then repeat this procedure for different $R$ to obtain $\sigma_\tau^2(R)$ as a function of $R$. These results are shown in Fig.~\ref{fig1}c. 

For a disordered non-hyperuniform packing, generated from a hyperuniform packing in which 10$\%$ of randomly chosen large and small particles are swapped, we observe the expected scaling $\sigma_\tau^2\propto (R/D)^{-2}$ (dashed line). By contrast, the variance of the hyperuniform packings exhibits a steeper decrease. A practical way to visualize hyperuniformity is to plot $(R/D)^2\sigma_\tau^2$ as a function of $R/D$ (Fig.~\ref{fig1}c). For the non-hyperuniform packing, this quantity remains constant. In a disordered jammed hyperuniform packing (i.e., the MRJ system) this quantity decreases with $R/D$ \cite{Zachary11}. Specifically, one expects the variance to exhibit the asymptotic scaling
\begin{eqnarray}
\sigma^2_{\tau}(R/D) \sim \left(\frac{D}{R}\right)^3[c_0 + c_1\ln R/D)] + {\cal O}\left(\left(\frac{D}{R}\right)^4\right)&   \label{eq:log} \\    
(R/D \rightarrow \infty), &\label{murks} \nonumber
\end{eqnarray}
where $c_0$ and $c_1$ are structure-dependent constants of order unity.

Fig.~\ref{fig1}d-f summarize the analysis is reciprocal space. Using reciprocal space relation (Eq.~\ref{FT}), we obtain direction-dependent spectral densities for both hyperuniform and non-hyperuniform packings, such as  the ones shown in Fig.~\ref{fig1}d\ and e. The origin, ${\bf k}=0$, is the center of the two power spectra, and the intensities vary radially as shown for one particular direction (indicated by the arrow in Fig.~\ref{fig1}d). Figure~\ref{fig1}d is obtained from a hyperuniform packing, and it exhibits a black spot at the center of the image. The ring around the black spot corresponds to nearest-neighbor location. i.e., to wavenumbers inversely related to the size of the colloids in the image. Figure~\ref{fig1}e is the power spectrum of the non-hyperuniform packing. As expected, it exhibits a white spot at the image center, indicating that the power spectrum does not vanish at very low wavevectors. Since the two power spectra are isotropic, they are readily angularly averaged to obtain the radial power spectra ${\tilde \chi}(k)$ in Fig.~\ref{fig1}f,
where $k=|\bf k|$ is the wavenumber. By contrast, one can very clearly see that the power spectrum of the hyperuniform packing is unlike that of the non-hyperuniform packing in that it vanishes as $k$ tends to zero.

These two methods will be utilized throughout the remainder of this paper, e.g., with experimental data. It is important to note that the methods and work discussed in this sub-section  employed ideal binarized images as materials for analysis. Therefore, these materials have no uncertainty in particle positions; furthermore, the present numerical data is not degraded by the discretization step that must be carried out when starting from actual experimental images. For example, when the system under investigation is a packed colloid of thermal micron-sized particles, optical microscopy cannot provide us with images of sufficient resolution to generate accurate binarized images of polydisperse samples. Specifically, optical aberration and scattering of light makes it difficult to precisely resolve the edges of each particle, and local variation of the image intensity on larger length scales prevents application of simple threshholding operations to generate binarized images. For these reasons, it is crucial to develop reliable procedures to reconstruct particle packings from the optical image data. Therefore, we will first explore simulated hyperuniform packings in order to assess the effect of the different reconstruction algorithms for assessment of hyperuniformity; these simulated packings will further enable assessment of the most important technical complications that can arise from experimental samples. After these steps we apply optimized methods for analysis of the experimental systems.

\section{Materials and Methods}

\subsection{Jammed disordered packings of PNIPAM particles}

The experimental samples were composed of soft poly(N-isopropyl acrylamide) (PNIPAM) microgel particles.  The samples were binary particle suspensions with PNIPAM particles of two different radii, $r_{l}\approx0.57$~\textmu m and $r_{s}\approx0.43$~\textmu m at $27\,^{\circ}{\rm C}$. The particles were synthesized by surfactant-free radical emulsion polymerization, as described elsewhere~\cite{still13}. A quasi-2D packing was prepared by confining the binary mixtures of large and small PNIPAM particles between two cover slips (Fisher Scientific) and then sealing the samle edges with optical glue (Norland 63)~\cite{Han2008}. 

\begin{figure}[t!b!]
\includegraphics[width=1.00\linewidth]{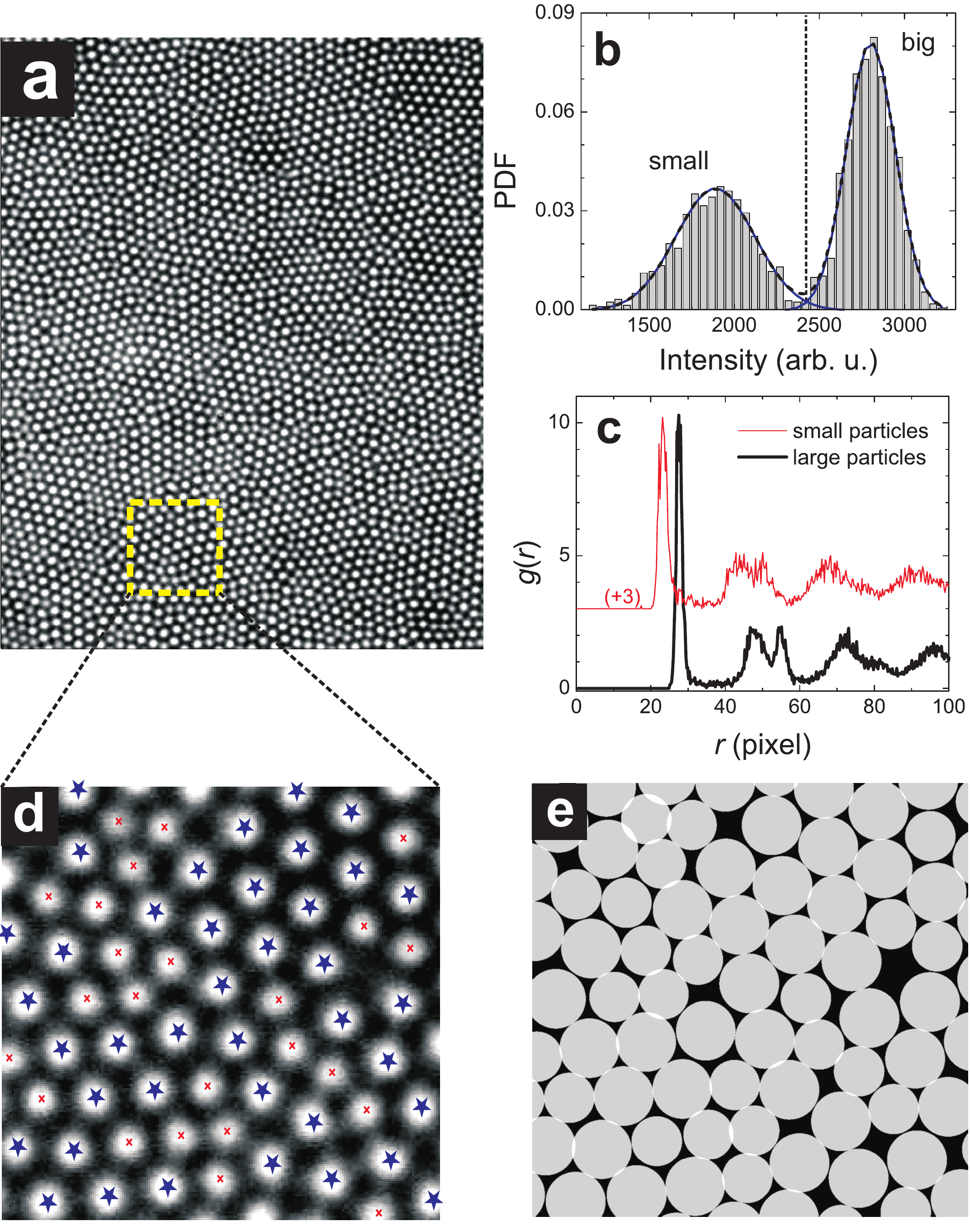}
\caption{(Color online) (a) Example of a raw (original) image showing a jammed packing of small and large PNIPAM spheres, each of which has a small degree of polydispersity from their associated mean. (b) Probability density function of individual particle intensity derived from particle tracking together with Gaussian fits for the two populations. The dashed line indicates the cutoff intensity for separation of small and 
large particles. (c) Pair correlation function for the small particles (red, thin) and the large particles (black, fat). The mean diameter of each population is inferred from the position of the first peak in the pair correlation function. (d) Snippet of the original image with the detected centers superimposed (red crosses: small particles; blue stars: big particles). (e) Reconstructed image using the PCF method. Note, on the scale of this figure, the reconstructions produced by the other two techniques described in the text, DLS and  j-PSR methods, are nearly indistinguishable. Therefore, they are not shown. The j-PSR method is comparatively superior for minimizing interparticle overlaps.  }
\label{fig2}
\end{figure}

Since PNIPAM is a temperature-sensitive polymer, the particle diameter could be controlled by changing temperature. Thus, the effective packing fraction of the sample was tuned {\it in-situ} by controlling the sample temperature with an objective heater (BiOptechs).  The average diameter of the large and small PNIPAM particles, and the corresponding packing fraction of the sample as a function of temperature can be found in the supporting information. Briefly, the temperature was set to $26\,^{\circ}{\rm C}$ so that the packing was near the jamming point. The trajectories of $N\approx4500$ particles in the field of view were extracted from a total of 3000 frames of video at 10 frames/s using standard centroid finding and particle tracking techniques ~\cite{Crocker1996}.  The time-averaged positions of the particles were used for image reconstruction, and the integrated intensity of each particle was used to divide spheres into two groups: large and small diameter (Fig.~\ref{fig2}a and b). 

\subsection{Generation of simulated hyperuniform packings}
\label{simulation}
We numerically generated 2D MRJ (maximally random jammed) packings of 200 to 10,000 hard repulsive particles using the algorithm developed by Zachary, Jiao and Torquato ~\cite{Zachary11}. This packing protocol is a classical molecular dynamics method in which initially small particles in  a unit cell under periodic boundary conditions undergo collisions with one another and  grow as a function of time at some expansion rate $\gamma$ until they jam. The expansion rate is chosen to be the fastest time consistent with strict jamming, which leads to MRJ states \cite{Do05d}. Each configuration consisted of a 75:25 binary mixture of large and small disks with size ratio of 1.33. 

We studied the effect of polydispersity index (PDI) up to PDI$=5\%$ for both large and small disks. We did so by generating such packings, wherein PDI is the standard deviation of the size distribution divided by their mean diameters. For each configuration, we started the packing algorithm with random positions of non-overlapping particles of desirable polydispersity but about 30\% smaller than their final sizes.   We then allowed the particles to grow at an expansion rate, $\gamma=dD/dt=10^{-2}D_0$, until a jammed state was reached; here $D$ and $D_0$ are, respectively, the final and initial disk diameter.

%---------------------------------------------------------------------------------------------------------------------------

\section{Packing reconstruction from images}
In order to reconstruct the entire packing with high accuracy, it is necessary to determine the center and  diameter of each colloidal particle. We devised three different reconstruction algorithms for this challenging task. The first approach is based on the sample’s pair correlation function (PCF), and we call it the ``PCF method." The second approach is based on the particle radius obtained from dynamic light scattering (DLS) experiments; we call it the ``DLS method''.  The third approach recovers the size of the particles by minimizing the overlaps and the gaps between particles, and we refer to it as the jammed-particle size-recovery method, i.e., the ``j-PSR method''.

We start the reconstruction from the optical microscopy images. The microscopy images are similar to the one in Fig.~\ref{fig2}a, which shows a typical packing of the binary PNIPAM colloids. Using the algorithm developed by \citeauthor{Crocker1996}~\cite{Crocker1996}, the centers of all the particles are readily identified. In our jammed packings, thermal motion is much smaller than the particle size, and we average over many frames to obtain highly accurate particle positions (centers). Next, an intensity based on the particle brightness is assigned to each detected particle; the histogram of intensity is plotted in Fig.~\ref{fig2}b. 
The histogram clearly exhibits two well-defined populations. These populations correspond to the small and large particles, and each individual particle is labeled as large or small, correspondingly (Fig.~\ref{fig2}d). The crucial next step is to assign a radius to each particle.

 \subsection{Reconstruction using pair correlation functions (PCF method)}
After the center of each colloidal particle has been identified and classified according to the population to which it belongs, the pair correlation functions, $g(r)$, of both the small and large colloids are computed. Here, by definition the quantity $\rho 2\pi r g(r) dr$ gives the expected number of particle centers in a annulus of differential thickness located at a radial distance $r$ from a typical particle center; $\rho$ is the number density. The pair correlation function results are shown in Fig.~\ref{fig2}c; the red curve is the pair correlation function of the small particles in the packing and the black curve is the pair correlation function of the large colloidal particles. The first peaks in the correlation functions give the average distance between nearest neighbors, and in a densely packed system like ours, we assume the mean diameter of each species to be identical to this distance.

Using the particles centers and the resultant mean diameters, the packing is reconstructed and is shown in Fig.~\ref{fig2}e. The white pixels that appear in the image correspond to zones wherein particle overlap occurs. Because polydispersity is not accounted for in this approach, overlaps between particles and the presence of rattler particles are unavoidable. Further, another possible source of error arises from the fact that the separation between large and small particles from Fig.~\ref{fig2}b is not perfect (albeit it is very good). Thus, errors may arise from the fact that some small particles are identified as large particles and \textit{vice versa}.

\subsection{Reconstruction using mean diameters obtained by dynamic light scattering (DLS method)}

In this approach, we define the particle diameters based on data from dynamic light scattering measurements. To this end, a standard DLS setup is employed to measure the mean particle sizes in diluted suspensions of both species at the experiment temperature. As was the case with the PCF method, the DLS method does not account for polydispersity, and the same limitations apply to both methods as a result. Note also, because DLS measures a hydrodynamic radius rather than a structural radius, the mean diameter of soft polymeric particles such as PNIPAM particles is likely to be slightly overestimated.

\begin{figure}
	\centering
		\includegraphics[width=1.00\linewidth]{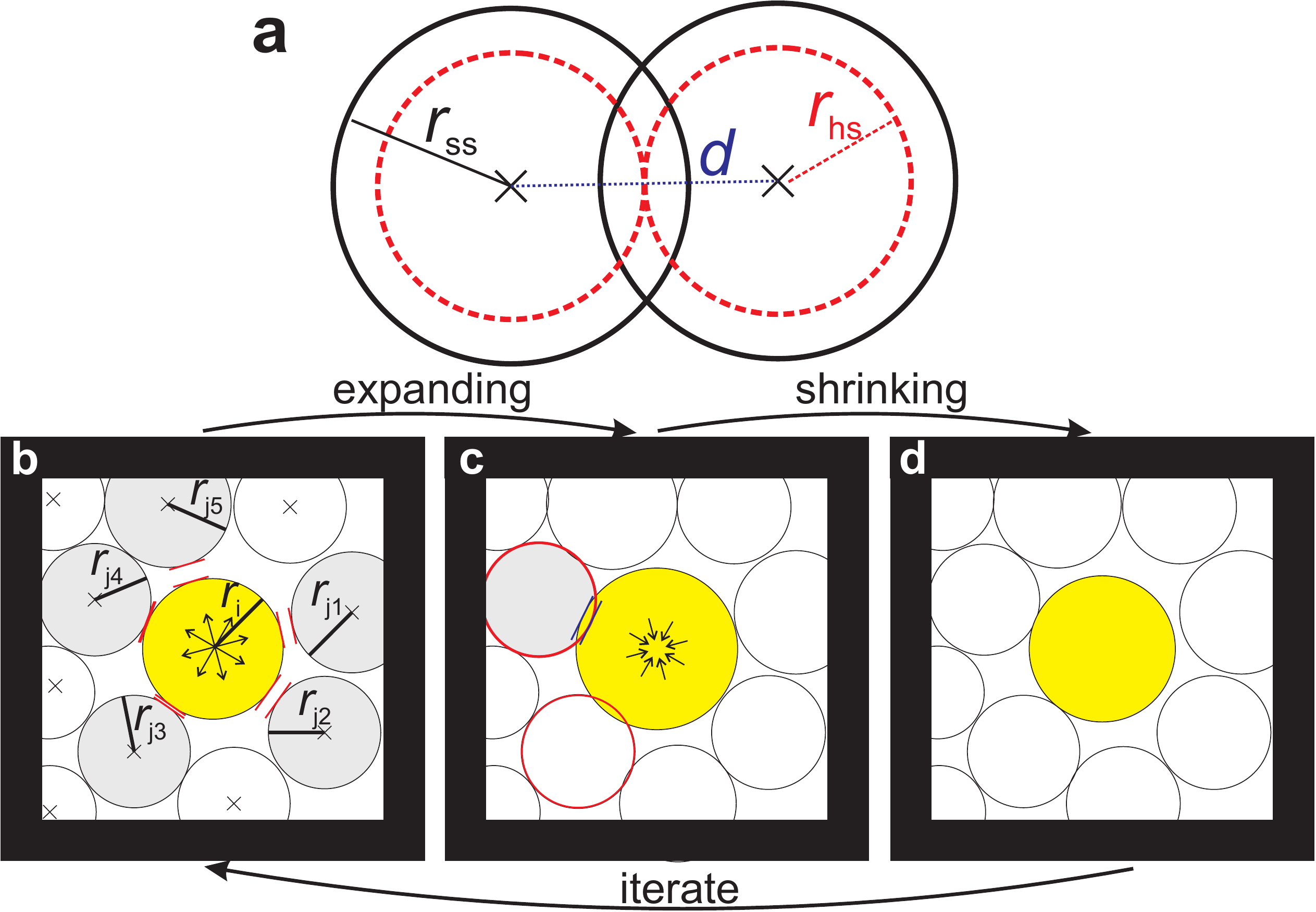}
	\caption{(a) Definition of ``true'' soft sphere radius $r_{ss}$ (solid circles) and hard sphere-like radius $r_{hs}$ (dashed circles). Soft spheres can overlap; centers and interparticle distance $d$ (dotted line) are unchanged in the hard sphere-like approximation, but particles touch only in one point.
	Cartoons describing the two cycles of our algorithm are shown in (b), (c) and (d); the same procedure is applied to all particles simultaneously, but here we highlight the process for the yellow particle. 
	(b) First step: Check which of the neighboring particles fulfill Eq. (\ref{eq:touch}) (i.e., grey particles); calculate average gap distance and expand circles according to Eq.~\ref{eq:add}. 
	(c) Determine which particle has the largest overlap with the yellow particle (e.g., one grey particle in this figure) and shrink particles according to Eq. (\ref{eq:sub}). 
(d) Iterate with step (a) using updated radii.}
	\label{fig3}
\end{figure}

\begin{figure*}[t!b!]
	\centering
		\includegraphics[width=1.00\linewidth]{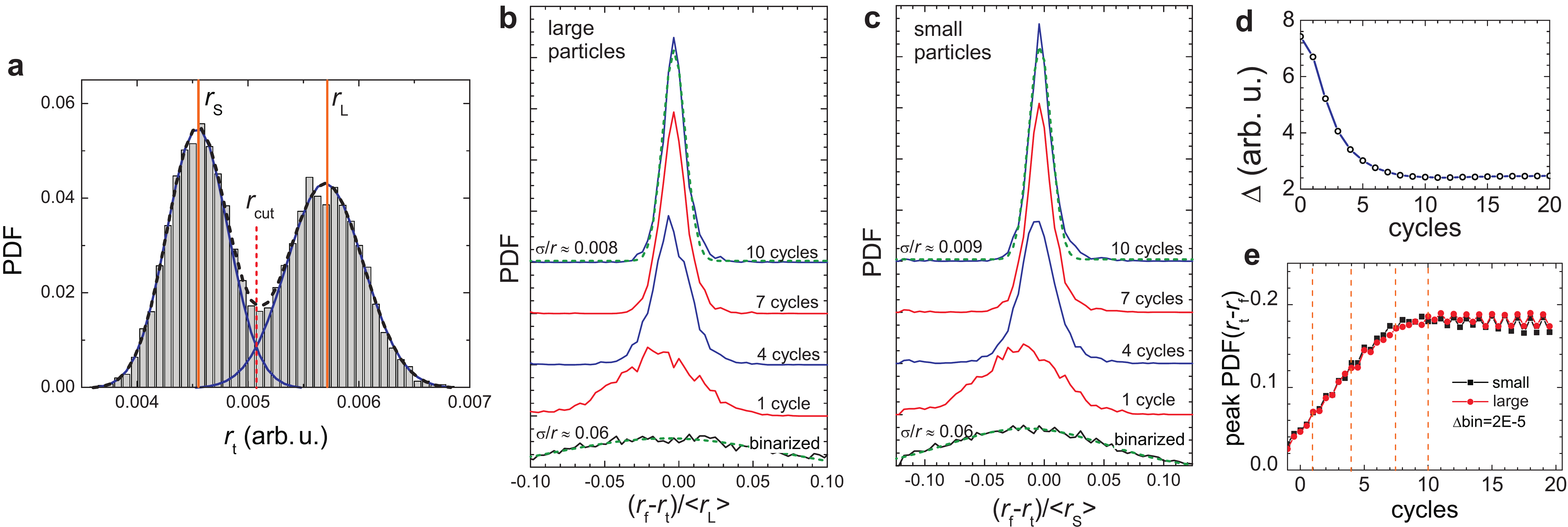}%{hyperuniform_hard.eps}
	\caption{(a) Particle size histogram with Gaussian fits of bi-polydisperse simulated hard sphere packing with $\phi\approx 0.84$ and PDI$\approx$0.06 for both sizes. To test the algorithm, we binarize the distribution by setting $r\equiv r_{s}$ for $r_{t}\le r_{cut}$ and $r\equiv r_{L}$ for $r_{t}> r_{cut}$. We then run the packing code using the following parameters: $p_{(+)}=p_{(-)}=0.4$, $p_{(-,ini)}=0.49$, and $\alpha=0.05$.
	(b) and (c) Evolution of fitted radii minus true radii after 1, 4, 7, and 10 full cycles compared to the initial binarized state for big and small particles, respectively. For the initial state and after 10 cycles, Gaussian fits are shown as well (dashed curves); distributions are about an order of magnitude narrower after 10 cycles compared to the initial PDI. 
	(d) The deviation parameter $\Delta$ (see text) decreases initially but is essentially constant after 10 cycles.
	(c) Fraction of particles in the most populated bin in the histograms of (b) and (c) (and all other cycles), a measure of ``distribution sharpness'' vs.\ cycles. The quality of the particle size fit is best after about 10 cycles and remains almost constant after that.
	Note that non-integer cycle numbers belong to the addition step, the initial (binarized) state is at cycle -0.5 and cycle 0 corresponds to the initial shrinking step before the first full cycle.
	}
	\label{fig4}
\end{figure*}

\subsection{Reconstruction using jammed-particle size-recovery (j-PSR) algorithm}

Our approach follows the spirit of Ref. \cite{Kurita11}. In this paper, an iterative algorithm was developed to measure the sizes of individual particles in two- and three-dimensional imaging experiments \cite{Kurita11}. This approach works best for hard spheres with well-known mean particle sizes and with positional data measured at different times.  

For soft particle systems, neighboring particles can be separated by gaps or they can overlap. Therefore, by contrast to hard-spheres, the diameter of the deformed sphere is somewhat ill-defined, as is the experimentally-determined mean radius of the soft spheres. Such is the case for the suspended hydrogel particles employed in the present experiments. 

Moreover, we typically investigate jammed packings in which particles do not move much over time, and therefore, contrary to the approach followed by Kurita \cite{Kurita11}, our analysis must rely on a single set of time-averaged data to derive particle positions. 

In order to deal with these issues, we developed a new algorithm that derives an estimate of the true particle radii starting from a single set of data about particle positions. In compressed packings of \textit{soft} particles, our algorithm measures the \textit{hard-sphere-like} particle size.  Specifically, we assign to each soft-sphere radius  $r_{ss}$  an equivalent  hard-sphere-like radius $r_{hs}$ that corresponds to the radius for which particles are just touching and not overlapping (Fig.~\ref{fig3}a). Thus, while the corresponding reconstruction of Fig.~\ref{fig2}d using the j-PSR method is almost indistinguishable from the reconstruction employing the PCF method (shown in Fig.~\ref{fig2}e), the former procedure is superior in that it  minimizes the degree of interparticle overlaps. This point will be further elucidated in Sec. \ref{detect}.

The main idea of j-PSR algorithm is to adjust individual particle size by minimizing both the overlaps and the gaps between nearby particles. The idea is illustrated graphically in Fig.~\ref{fig3}b-d. Each cycle of the fitting procedure consists of two steps: the first step is a small expansion of particle sizes to minimize gaps between neighboring particles. Then, in the second step, the particles are decreased in size to avoid overlap. 
We let $d_{ij}$ denote the center-to-center distance between any two particles labelled by $i$ and $j$; this distance is usually known almost exactly. Let the approximate particle radii before the $n$-th step be denoted by $r_i^{n-1}$ and $r_j^{n-1}$, respectively. In step one, the gaps between particles are corrected using the following equation:
\begin{equation}
r_i^n=r_i^{n-1}	+ p_{(+)}\frac{\sum\limits_{j=1,m}\left(d_{ij}-r_i^{n-1}-r_j^{n-1}\right)}{m}.
\label{eq:add}
\end{equation}

Here $j$ runs over $m$ nearest-neighbors of particle $i$ for which 
\begin{equation}
	0\le d_{ij}-r_i-r_j\le \alpha \left(r_i+r_j\right).
	\label{eq:touch}
\end{equation}

Equation (\ref{eq:add}) implies that each particle is to be expanded by a fraction, $p_{(+)}$, of the mean gap between itself and its nearest-neighbors. To utilize this procedure, we must define nearest-neighbor particles. A number of definitions can be used, but here we call particles $i$ and $j$ nearest-neighbors, if the gap between $i$ and $j$ is not larger than a fraction, $\alpha$, of the sum of their radii $r_i^{n-1}+r_j^{n-1}$. This criterion defines the touching neighbors in a more constrained manner; they are not necessarily the geometric neighbors as determined by a Voronoi construction. In practice we perform this expansion for all particles simultaneously which, in turn, will lead to some overlaps between neighbors.

Thus, the second step in each cycle consists in shrinking particles (i.e., decreasing particle diameter) to minimize the overlap. 
For this purpose, we recompute a new radius for each particle using the following equation:
\begin{equation}
r_i^{n+1}=r_i^n-p_{(-)}\left(r_i^n+r_j^n-d_{ij}\right)_{max}.	
\label{eq:sub}
\end{equation}
Here $\left(r_i^n+r_j^n-d_{ij}\right)_{max}$ is the largest absolute overlap between particle $i$ and any of its neighbors, $j$. 
Note that setting the parameter $p_{(-)}=0.5$ removes all overlap, but by setting it to be a little smaller than $0.5$ we can usually obtain better results, since we then avoid over-corrections due to a few particles with grossly overestimated radii. Thus, every cycle consists of the two consecutive steps described by Eqs.~\ref{eq:add} and \ref{eq:sub}. In practice it has proven useful to include an initial shrinking step before the first cycle, replacing $p_{(-)}$ by $p_{(-,ini)}\approx0.5$ in order to start all cycles from an overlap-free configuration.

A useful way to follow the system evolution as a function of cycle (iteration) number is to calculate a so-called ``deviation parameter" $\Delta$.  $\Delta$ is the sum of the absolute values of gaps and overlaps between neighbors for the entire packing, i.e.,
\begin{equation}
\Delta=\sum\limits_{i}\sum\limits_{j}\left|d_{ij}-r_i-r_j\right|.
\end{equation}
Note, the iterative procedure minimizes $\Delta$, but $\Delta$ should remain finite even if the fit is perfect, because not all neighbors actually touch. 

We tested the algorithm by applying it to simulated packings of particles generated using the procedures described in Section~\ref{simulation}. The size distribution of an exemplary packing is shown in Fig.~\ref{fig4}a.  In this case, we assume a bi-polydisperse packing with a size ratio of $\approx1.3$ and a polydispersity index of $PDI=0.06$ for both large and small particles. The size ratio is similar to the experimental value, and we tested our algorithm for a wide range of polydispersities. The first step is to distinguish between small particles (with mean radius $r_{s}$) and large particles (with mean radius $r_{L}$). 
We then binarize the distribution: particles of radius below a certain threshold value $r_{cut}$ are considered as small and their radius is $r_s$, particles of radius above $r_{cut}$  are considered as large and their radius is $r_L$. By doing such a radius assignment, we pretend that we can only distinguish large and small particles (like in the microscopy experiment). From this starting point, we try now to recover the true particle sizes (Fig.~\ref{fig4}a.)
Starting from these binarized radii, we then run the algorithm using the following parameters: $p_{(+)}=p_{(-)}=0.4$, $p_{(-,ini)}=0.49$, and $\alpha=0.05$.
At each cycle, we can compare the recovered fitted radii $r_f$ to the true values, $r_t$, given by the simulations. We plot the evolution of probability distribution functions of $r_{f}-r_{t}$ after 1, 4, 7, and 10 cycles in direct comparison to the starting probability distribution where particles are just binarized for large and small particles (Fig.~\ref{fig4}b,c).

A perfect recovery of the true particle radii would correspond to a delta function at $r_{f}-r_{t}=0$. Evidently, after 10 cycles, we recover the true radii with very high accuracy; the top curves in Fig.~\ref{fig4}b\&c can be fitted by Gaussian curves much narrower than the binarized curve with polydispersity, $PDI=\sigma/r\approx0.06$.
After 10 cycles, $\sigma/r<0.01$, which is comparable to the results by Kurita et al. \cite{Kurita11} obtained after time averaging. 

%---------------------------------------------------------------------------------------------------------------------------

\section{Hyperuniformity from reconstructions based on simulated packings}
\label{detect}

\subsection{Finite sample-size effects} 
\begin{figure}
\includegraphics[width=1.00\linewidth]{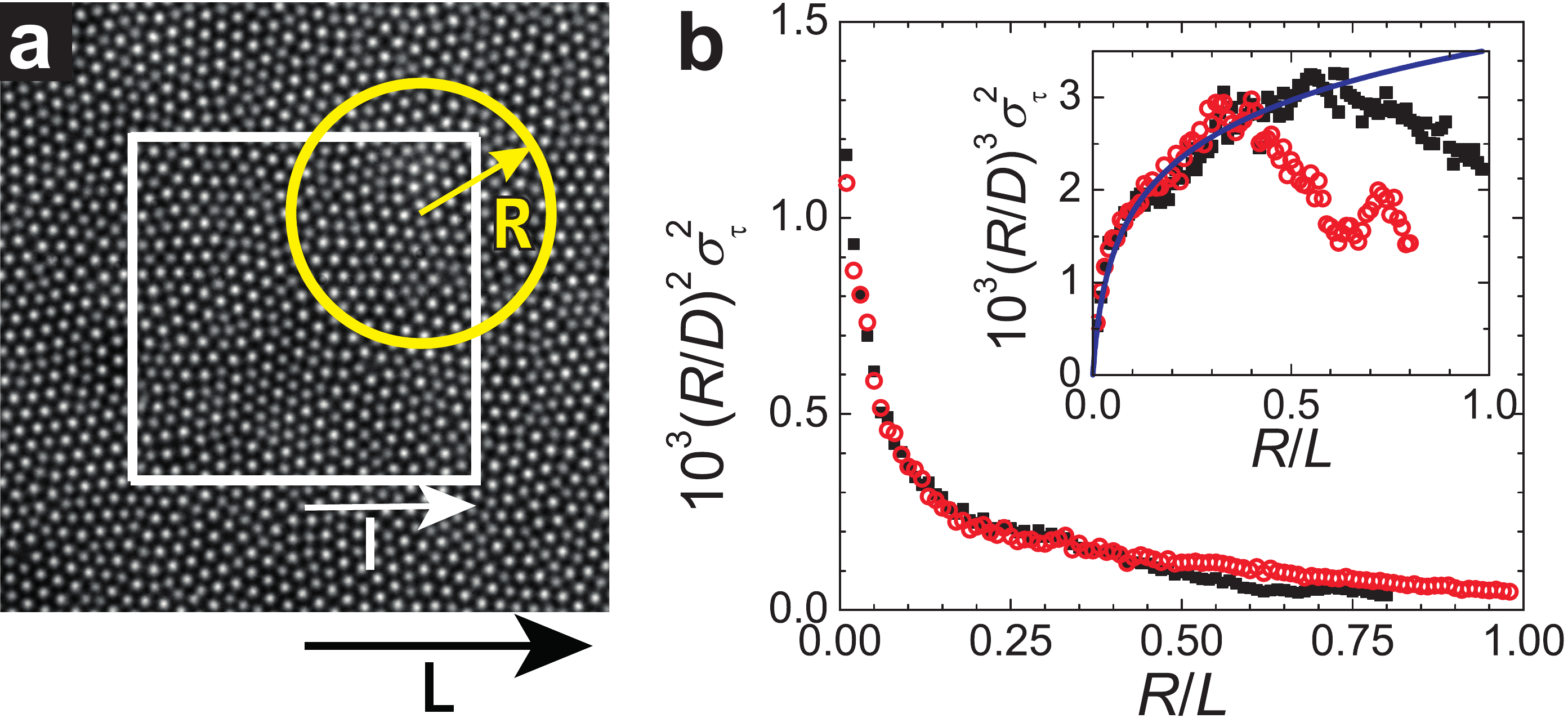}
\caption{ (color online)  
(a) Definitions of the size parameters. The square box has a side of size $2L$, the disk of inspection has radius $R$, and its center can span a box of size $l$. The hyperuniform jammed system in the box was simulated under periodic boundary conditions. 
(b) $\left(R/D\right)^2\sigma_\tau^2$ as a function of $R/L$ for the box of size $2L$ (open circles), and for the larger system containing 9 stitched images with box side length $6L$ (solid squares). The signal decreases as expected for a hyperuniform system. Inset: $\left(R/D\right)^3\sigma_\tau^2$ as a function of $R/L$. The drop-off of signal for $R/L=0.35$ (open circles) and $R/L=0.70$ (solid squares) is a finite sample-size effect. }
 \label{fig5}
\end{figure}

The hyperuniformity concept was developed for infinitely large systems. Of course, when working on real experimental systems, one has to account for the finite sample-size. Here we employ MRJ packings of spheres known to be hyperuniform and generated by numerical techniques in order to assess finite sample-size effects. One such hyperuniform simulated packing was generated, and we explore the system comprised within a square box with a side-length of $2L$ (Fig.~\ref{fig5}a). (Note, the simulated packing assumed periodic boundary conditions.) To detect the hyperuniformity of the packing, we use both the direct-space approach for the local-volume fraction variance, Eq.~\ref{tau}, and spectral (reciprocal-space) method via Eq.~\ref{FT}. 

Within the simulation box, we randomly place circular windows of radius $R$.  
Because the entire circular window is constrained to be included in the box, the window centers are located inside a smaller square box with side-length of $2l$. The quantities $L$, $l$ and $R$ are chosen such that $l+R \le L$. In Fig.~\ref{fig5}b, we plot $\left(R/D\right)^2\sigma_\tau^2$ as a function of $R/L$. 
As mentioned earlier, for a non-hyperuniform system, $\left(R/D\right)^2\sigma_\tau^2$ is constant with $R/L$ for large $R$ whereas, for a hyperuniform system, $\left(R/D\right)^2\sigma_\tau^2$ exhibits a large decrease as a function of $R/L$, as described by the logarithmic asymptotic relation in Eq.~\ref{eq:log}.

The results for the finite system with a side-length of $2L$ correspond to the red open-circles in Fig.~\ref{fig5}b. We see that for $R/L$ between $0$ and $0.35$,  $\left(R/D\right)^2\sigma_\tau^2$ is not constant, and it is close to the $\ln (R)/R$ law expected for hyperuniform packing of spheres \cite{Zachary11}.  In the inset of Fig.~\ref{fig5}b., we plot  $\left(R/D\right)^3\sigma_\tau^2$ as a function of $R/L$. For a hyperuniform system, this quantity should increase as $\ln R$.  At $R/L\gtrapprox0.35$, the data (derived from simulation) shows a strong and sharp decrease followed by oscillations. To assess the origin of this behavior, we construct a sample that is larger in size.  Since the original packing was simulated under boundary periodic conditions, we can construct a larger fictitious system of size $6L$ x $6L$ by simply stitching the original sample nine times. The hyperuniformity of the larger system was similarly determined and is represented by the black solid-squares in Fig.~\ref{fig5}b. 

Interestingly, we can see that the red and black data points superimpose for $R/L$ ranging from $0$ to $0.35$. For larger values of $R/L$, however, the signal for the stitched system continues to follows a logarithm law until it finally decreases for $R/L\gtrapprox0.7$. 

These observations suggest that the drop-off observed for the red data points above $0.35$ is indeed due to the finite-size of the box. For this reason, we will apply the criterion that $R/L \le 0.35$ to detect hyperuniformity from measurements.

\subsection{Comparison of reconstruction methods based on simulated packings}
In this section we examine which of the three reconstruction methods is most effective for hyperuniformity analysis of colloidal packings similar to our experimental data. Importantly, we take the finite sample-size effects discussed in the last sub-section into consideration for these comparisons. Specifically, we generate hyperuniform packings made of two populations of hard spheres, i.e., particle configurations with no interparticle overlaps. We study populations of particles which have a polydispersity of $0\%$, $3\%$ and $5\%$, respectively. Starting with perfectly known and well-defined particle configurations, we reconstruct the packing using the PCF, DLS and j-PSR techniques described earlier.

Interestingly, all reconstructions obtained from these three methods yield very similar overall packing fraction. However, both the PCF and the DLS methods deliver significant overlap between particles, i.e., of order 1\% of the overall packing fraction. By contrast, the amount of overlap produced by the j-PSR algorithm depends on the $p_{(-)}$-parameter in the last subtraction step, and the resulting overlap is usually much smaller than the PCF and DLS methods (e.g., it is zero for $p_{(-)}=0.5$). Furthermore, and importantly, the j-PSR reconstruction reliably reproduces the true polydispersity in the particle distributions.

\begin{figure}
\includegraphics[width=1.00\linewidth]{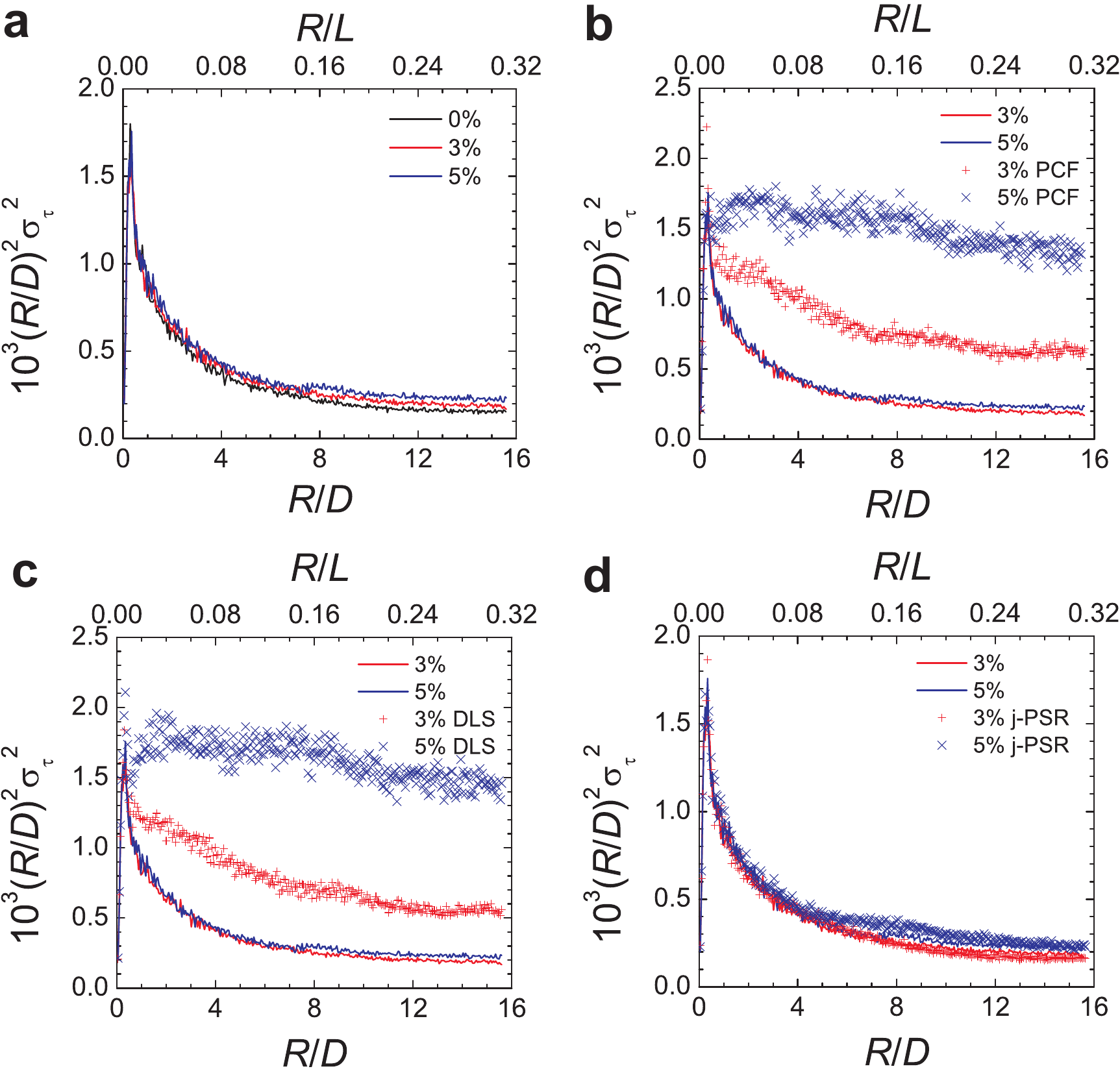}
\caption{ (color online)  Direct-space measurements of hyperuniformity for simulated packings of MRJ systems made of two polydisperse particle populations. (a)  Here, $p=0\%$ (black), $p=3\%$ (red) and $p=5\%$ (blue). Each generated packing shows a hyperuniformity signature. (b) Here, $p=3\%$ (red) and $p=5\%$ (blue) and the reconstructed packing using the PCF method $p=3\%$ (red cross) and $p=5\%$ (blue cross)  (c) Here, $p=3\%$ (red) and $p=5\%$ (blue) and the reconstructed packing using the DLS method $p=3\%$ (red cross) and $p=5\%$ (blue cross)  (d) Here, $p=3\%$ (red) and $p=5\%$ (blue) and the reconstructed packing using the j-PSR method $p=3\%$ (red cross) and $p=5\%$ (blue cross) }
 \label{fig6}
\end{figure}

\begin{figure}
\includegraphics[width=1.00\linewidth]{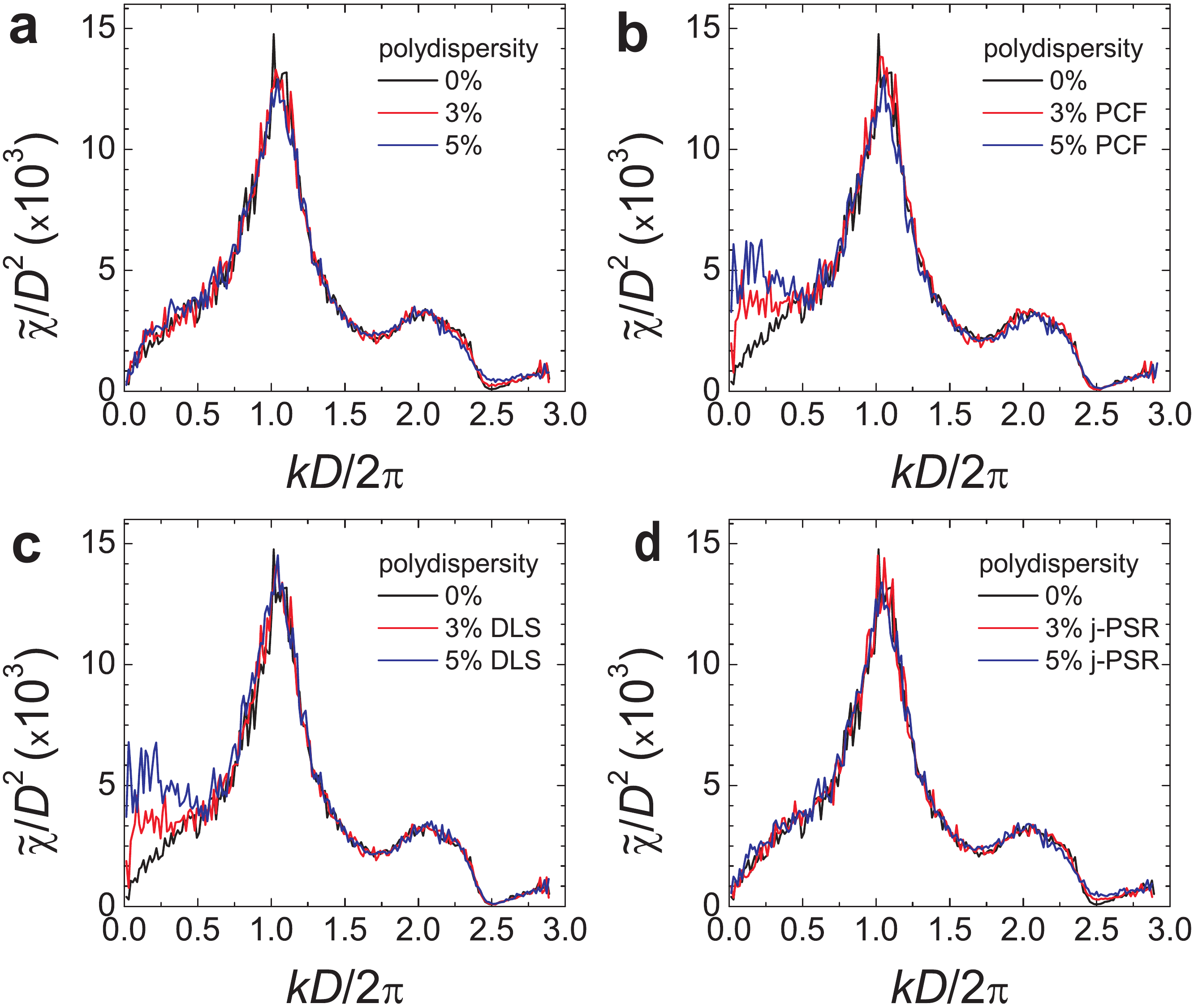}
\caption{ (color online)  Reciprocal-space measurements of hyperuniformity for simulated packings of MRJ systems made of two polydisperse particle populations. (a)  Here, $p=0\%$ (black),$p=3\%$ (red) and $p=5\%$ (blue). Each generated packing shows a hyperuniformity signature. (b) Here, $p=0\%$ (black) and the reconstructed packing using the PCF method $p=3\%$ (red ) and $p=5\%$ (blue)  (c) Real simulated packing $p=0\%$ (black)  and the reconstructed packing using the DLS method $p=3\%$ (red) and $p=5\%$ (blue)  (d) Here, $p=0\%$ (black) and the reconstructed packing using the j-PSR method $p=3\%$ (red) and $p=5\%$ (blue)}
 \label{fig7}
\end{figure}

Next, we determine which (if any) reconstructed samples exhibit signatures of hyperuniformity. Figure ~\ref{fig6}a shows the results in direct space for three simulated packings; in these packings each population of small and large particles have a polydispersity of $0\%$, $3\%$ and $5\%$, respectively. In this figure, $(R/D)^2\sigma_\tau^2$ is plotted as a function of $R/D$ and $R/L$. Importantly, with knowledge of the correct size of each individual particle, the simulated packings exhibit very similar trends. In particular, $(R/D)^2\sigma_\tau^2$ does not remain constant with the normalized window radius: the packings generated by the simulation are hyperuniform. 

We can readily assess how the packing reconstructions derived via different methodologies (PCF, DLS, j-PSR) affect the diagnosis of hyperuniformity. Figure~\ref{fig6}b compares the packing obtained by the PCF method for two polydisperse populations of colloids against the original simulated packings. For both cases we can see in Fig.~\ref{fig6}b that the curves obtained from reconstruction strongly differ from those obtained from the simulated packings. \textit{This observation suggests that neglecting the polydispersity of each population of colloids, and replacing their diameters by the mean diameters obtained using the pair correlation functions, produces small errors in the packing reconstruction that strongly affect the diagnosis of hyperuniformity.}  We have found that making such an assumption for any sample with polydispersity higher than $3\%$ results in a \textit{false negative} diagnostic. The same conclusions are drawn from the results presented in Fig.~\ref{fig6}c, wherein the DLS method is used for reconstruction. 

By contrast, Fig.~\ref{fig6}d shows the same curves for the packing obtained using the j-PSR technique. Though the packing obtained by this technique still introduces errors, the curves obtained are very similar to those obtained from the original simulated packings. The reconstructed packings show clear signatures of hyperuniformity. To conclude from this study, we see that when the degree of polydispersity is of order $3\%$ or larger, then any reconstruction technique that assumes monodisperse particles results in the false diagnosis, i.e.,  a true hyperuniform packing does not appear to be hyperuniform when not accounting for particle polydispersity distributions. 

The same conclusions can be drawn by investigating the image power spectrum in the reciprocal space via (Eq.~\ref{FT}). Figure~\ref{fig7}a shows the power spectra for simulated packings for which both populations have a polydispersity of $0\%$, $3\%$ and $5\%$, respectively. The spectral densities or power spectra $\tilde{\chi}$ of these images are almost indistinguishable, and indeed they vanish at very low wavenumbers. However, the power spectra associated with the reconstructed images using either the PCF or DLS methods do not overlap with the power spectrum obtained from a hyperuniform packing (Fig.~\ref{fig7}b and c)at low wavenumbers. The spectral densities clearly do not vanish. Thus, for a $3\%$ polydisperse sample, reciprocal-space measurements show that the system is not hyperuniform. Finally, Fig.~\ref{fig7} d shows that the power spectra obtained from packings reconstructed using the j-PSR technique indeed vanish in the long-wavelength limit, which is consistent with the results obtained from direct space measurements.

Overall, the direct-space methods appear to be superior to the reciprocal-space method for detection of hyperuniformity when the system is indeed hyperuniform. Some theoretical arguments shed light on this question. It was observed in Ref. \cite{Torquato03} that the window size need only be about an order of magnitude larger than the nearest-neighbor spacing $D$ in a point pattern to estimate the long-range scaling of the number variance $\sigma^2_N(R)$ [Eq. (\ref{N})] with $R$. This is due to the fact that that the corrections to the leading order hyperuniform term in the asymptotic expansion for large $R$ are  comparatively negligible to this dominant contribution for such moderately-sized windows \cite{Torquato03}. On the other hand, in order to assess hyperuniformity through the spectral condition (\ref{S}), one must have very large systems to determine accurately the long-wavelength limit, i.e., wavenumbers tending to zero. The sizes of the relatively small windows in the direct-space procedure can be appreciably smaller than the size of the large system needed to get accurate results for the structure factor $S(k)$ in reciprocal space.

An analogous analysis applies to the local-volume fraction variance $\sigma^2_{\tau}(R)$. To see this explicitly, consider the asymptotic relation (\ref{eq:log}) that applies to 2D disordered hyperuniform MRJ packings \cite{Zachary11}. Recall that the leading-order nonhyperuniform term of order  $(D/R)^2$ is identically zero or, in practice very close to zero. (If this were not the case, then the scaling would be dominated by this leading order term and hyperuniformity would not even be a consideration.)  Since the constants $c_0$ and $c_1$ in Eq. (\ref{eq:log}) are of order unity, and because the implied constant multiplying the correction of order $(D/R)^4$ is also of order unity, it is clear than when $R/D$ is a relatively small number that the hyperuniformity terms will dominate. This effect is to be contrasted with the spectral condition (\ref{chi}), which requires very large systems to access the small-wavenumber behavior accurately.

\begin{figure}
\includegraphics[width=1.00\linewidth]{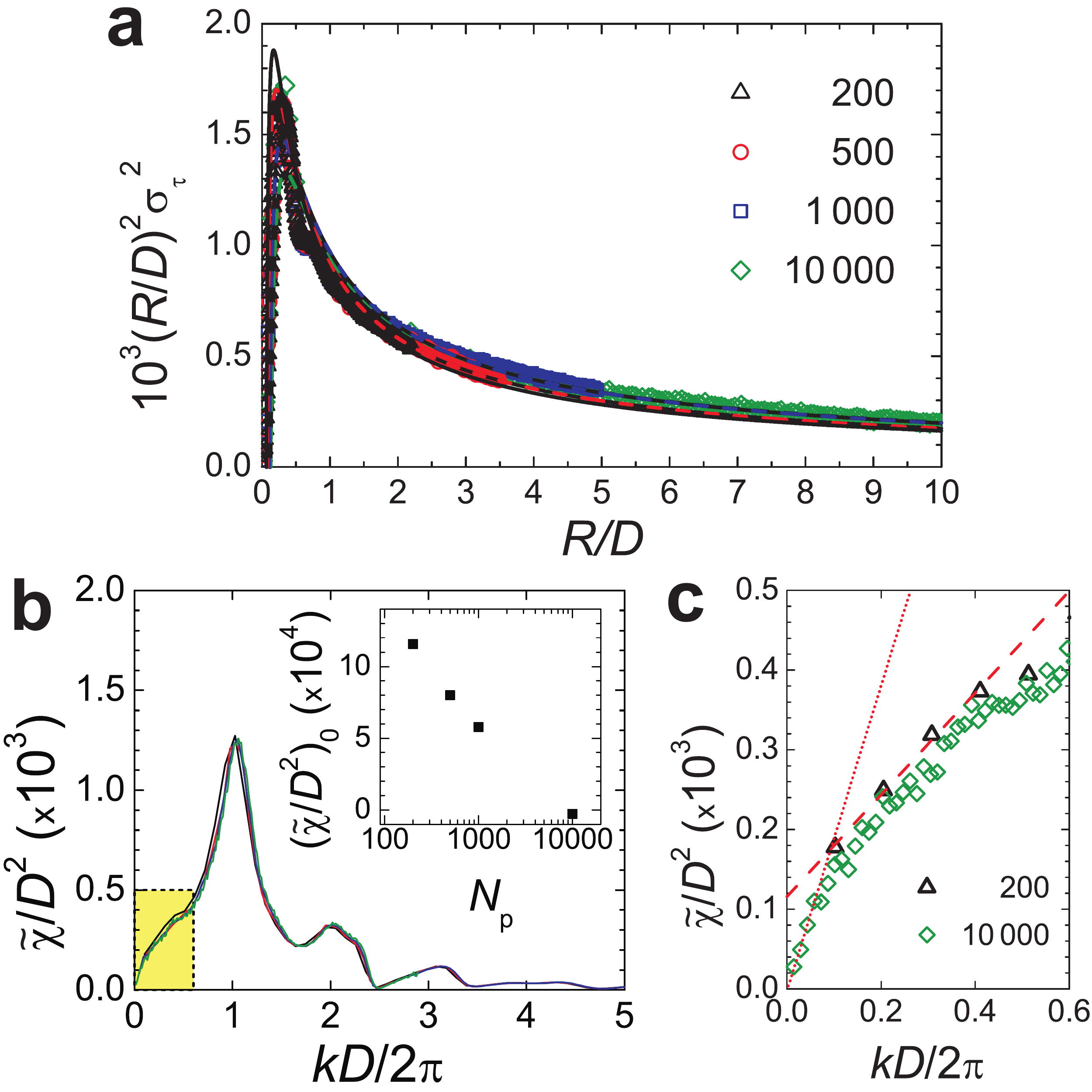}
\caption{(Color online) \textbf{a} Results for the hyperuniformity test performed in the direct space for simulated packings consisting of 200 (black triangle), 500 (red circle), 1,000 (blue square), and 10,000 (green diamond) particles.  Dashed lines are fitted curves with Eq.~\ref{eq:log}. \textbf{b} Results for the hyperuniformity tests performed by computing the power spectrum of the reconstructed images for the same simulated packings.  \textbf{c} magnified view for 200- and 10,000-particle packings in small $kD/2\pi$ region.  Red dashed lines for the fit to the first four data points for each system size.  Intersections of linear fits with the vertical axis, ${\tilde \chi}/D^2(k=0)$, for all four system sizes are plotted in the insert in \textbf{b}.}
 \label{size_effect}
\end{figure}

To demonstrate these theoretical arguments more explicitly, we analyze the effect of system size with simulated packings consisting of different numbers of bidisperse particles, as shown Fig.~\ref{size_effect}, and compare the direct-space and reciprocal-space measurements. The curves for both direct-space and reciprocal-space measurements are averaged from 100, 20, 10, and 10 independent configurations of 200-, 500-, 1,000-, and 10,000-particle packings, respectively. We average over configurations to remove any variations due to configurational fluctuations, thereby allowing us to more accurately assess system-size effects. 

For the direct-space measurements, shown in Fig.~\ref{size_effect}a, despite limited system size, even curves for 200- and 500-particle packings are fitted well with Eq.~\ref{eq:log}, leading to a conclusion of hyperuniform configuration. The extrapolated part from Eq.~\ref{eq:log} for small systems agrees well with the curve for the largest system with 10,000 particles.  Particularly, the fitted curve of 1,000-particle packings is indistinguishable with that of 10,000-particle packings. By contrast, for reciprocal-space measurements shown in Fig.~\ref{size_effect}b, even though the overall power spectra are the same for different system sizes, the small systems limit access to small $k$ values, as shown in the insert in Fig.~\ref{size_effect}c. We fit the first four data points from the smallest value of $kD/2\pi$ for each system size and extrapolate to the origin, as shown in Fig.~\ref{size_effect}c for 200- and 10,000-particle packings. While the extrapolated line for 10,000-particle packings goes nicely to the origin, the one for 200-particle packings shows a clear positive intercept with the vertical axis, suggesting non-hyperuniform packings. Those vertical intercepts, ${\tilde \chi}/D^2(k=0)$, versus system size, $N_p$, are plotted as the insert in Fig.~\ref{size_effect}b. Clearly, reciprocal-space measurements tend to result in \textit{false negatives} in diagnosing hyperuniformity for samples with limited size. 

The different effects of system size on direct-space and reciprocal-space measurements suggest that direct-space measurements are superior for assessment of hyperuniformity in systems of limited size which are often encountered in experiments. Moreover, when large systems cannot be realized, a more accurate determination of hyperuniformity can be achieved by averaging many similar configurations of relatively small size using direct-space measurements.

\section{Hyperuniformity in experimental jammed packings of soft spheres}

Guided by the lessons learned from simulated packings, we finally test the three reconstruction procedures on an experimental jammed (and disordered) packing of PNIPAM particles. The samples were made using procedures described above. The packings are reconstructed using the three methods, DLS, PCF and j-PSR, and then the hyperuniformity of the reconstructed packings is tested using both direct space and the reciprocal space approaches, as describe in Sec. \ref{detect}.

\begin{figure}
\includegraphics[width=0.7\linewidth]{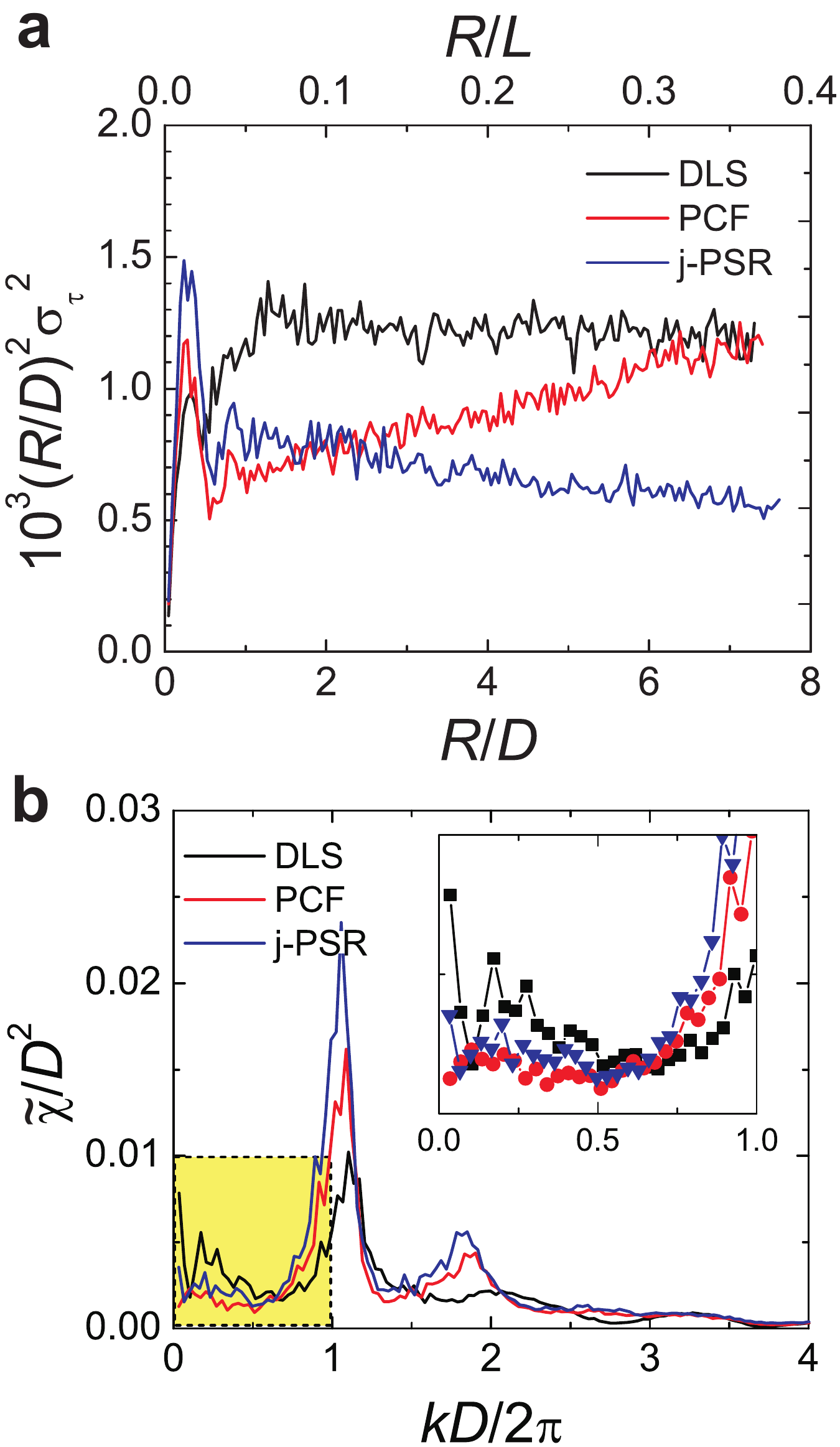}
\caption{(Color online) \textbf{a} Results for the hyperuniformity test performed in the direct space. Signature of hyperuniformity is detected when performed on a packing reconstructed by the j-PSR method. Packings reconstructed using DLS or PCF method do not show signs of hyperuniformity. \textbf{b} Results for the hyperuniformity tests via the power spectrum of the reconstructed images. The power spectra do not completely vanish at very low wavelengths.}
 \label{res_poly_packings}
\end{figure}

The results are plotted in Fig.~\ref{res_poly_packings}. In direct space, measurements of the experimental packing confirm observations made on simulated packings.
Specifically, the PCF and DLS methods suggest that the PNIPAM packing is not hyperuniform. Using the j-PSR method, however, the polydispersity index is found to be 6$\%$ and 3$\%$ for small and large particles, respectively. Furthermore, the packing obtained using the j-PSR method, which includes the effect of polydispersity, strongly suggests that the sample is hyperuniform. Notice that the scaled volume-fraction fluctuations deviate from a linear law, as can be seen in Fig.~\ref{res_poly_packings}a. Not surprisingly, hyperuniformity is not confirmed by the reciprocal-space measurements. None of the three power spectra obtained from the three methods vanish at very long wavelengths, although the signals for the j-PSR and the PCF methods plateau at a very low wavenumber close to $2.5\times10^{-3}$. These observations are consistent with the simulations studies that clearly suggest the direct-space approach is superior for detection of hyperuniformity (e.g., in samples of finite-size). Taken together, the results strongly suggest that the 2D jammed PNIPAM packing is hyperuniform, and the observations and analysis provide a framework to decipher whether these types of many-particle systems are truly hyperuniform or ``nearly" hyperuniform or are not hyperuniform \cite{He13,xie13}.

\section{Conclusions}
To conclude, we have devised a general methodology that facilitates diagnosis of hyperuniform particle packings from typical microscope images of colloidal packings.  In finding an optimized approach, we have investigated three different reconstruction methods that can be employed with experimental microscopy images. Studies with hyperuniform simulated data enabled us to assess the validity of the methods, and to understand the complications of particle polydispersity.  Importantly, we found that any procedure that neglects polydispersity as small as $5\%$ fails to reconstruct the original packing with an accuracy that is high enough to determine the hyperuniformity of the packing.  This observation is consistent with previous work reported by Berthier \textit{et al.}\cite{Berthier11}.   In order to properly account for polydispersity, we developed a new algorithm, the so-called j-PSR algorithm, which yields reconstructions that are accurate enough to decipher hyperuniformity. In addition, we discovered and explained why direct-space procedures that determine whether a packing is hyperuniform are more accurate than their reciprocal-space counterparts, i.e., the differences are due to the size of the finite system that can be accessed experimentally. Finally, using the results of our analysis we are able to prove that our 2D jammed and disordered systems composed of soft PNipam particles are indeed hyperuniform.

\begin{acknowledgments}
R.~D.\ and L.~A.~H.\ were supported by CNRS and Solvay. S.~T.\ was supported by the National Science Foundation under Grants DMR- 0820341 and No. DMS-1211087. Y.~X., T.~S., and A.~G.~Y.\ were supported by the National Science Foundation under Grants DMR12-05463, DMR-1305199, PENN MRSEC DMR11-20901, NASA NNX08AO0G. {\it Author contributions:} R.D., Y.E. initiated the project, performed the experiments, analyzed the data and wrote the article, T.S. developed the j-PSR algorithm, analyzed the data and wrote the article , L.A.H. supervised the research, A.G.Y and S.T. analyzed the data,  supervised the research and wrote the article.
\end{acknowledgments}

% Create the reference section using BibTeX:

%\bibliography{refs}
%

%
% ****** End of file template.aps ******
\end{document}